# HB morphology and age indicators for metal-poor stellar systems with age in the range of 1 to 20 Gyr

F. Caputo[1], S. Degl'Innocenti[2]

[1]Istituto di Astrofisica Spaziale, CNR, C.P. 67, I-0044 Frascati
[2]Dipartimento di Fisica, Universita' di Ferrara, Via Paradiso 12, I-44100 Ferrara

## Summary

Isochrone computations and horizontal branch (HB) models for $Y(MS)=0.23$ and two values of $Z$ (0.0001, 0.0004) are used to derive constraints on the age indicators and HB morphology of metal-poor clusters with age $t$ (in Gyrs) in the range of $1\leq t \leq 20$. It is found that for fixed metallicity the luminosity $L(3.83)$ of the zero age horizontal branch (ZAHB) at the RR Lyrae gap depends on age ($\log L(3.83)=$const$-0.205/t$), and that the difference in visual magnitude between main sequence turn-off (TO) and ZAHB may be calibrated in terms of age and metallicity as $\log t=-0.082-0.057(\log Z+4)+0.374\Delta V$(TO-ZAHB). The results of synthetic HB computations show that at fixed metallicity the mean luminosity of RR Lyraes depends on HB morphology, as well as that the mean luminosity of purely red horizontal branches depends on the age and amount of mass loss during the red giant phase. Such features should be properly taken into account when the cluster age is derived from the observed difference in magnitude between main sequence TO and actual horizontal branch (rather than ZAHB). Moreover, it is shown that the observed HB morphology provides us with an easy estimate of the upper limit of the cluster age. Since the age upper limit increases when increasing $Z$, while the absolute age which are derived from the difference $\Delta V$(TO-HB) mildly decreases, some useful constraints to the cluster metallicity could be also derived.

Key words: stars: evolution - stars: Population II - Galaxy: clusters: general - Galaxies: stellar content

# 1. Introduction

Constraining the age of stellar systems is among the more interesting opportunities provided by stellar evolution theory. For this reason a great deal of efforts has been devoted to derive theoretical constraints to the age of galactic clusters and, in particular, of old globulars membering the Galactic halo. In this case, a well established method relies on the observed difference in visual magnitude $\Delta V$(TO-HB) between the horizontal branch at the RR Lyrae gap (as observed or extrapolated in the absence of variables) and the main sequence (MS) turn-off (TO), and several calibration of such a parameter as a function of the overall metallicity $Z$, original helium content $Y$(MS), and age ($t$ in Gyr) may be found in the literature of the last decades.

In its original form the procedure relies on the assumption that the mean luminosity of RR Lyrae variables $L$(RR) (in solar unit) depends only on the cluster chemical composition, being independent of both the cluster age and HB type, i.e. the observed distribution among blue (B), variable (V), and red (R) HB stars, as given (Lee 1989) from the ratio (B-R)/(B+V+R). However, synthetic HB models (see e.g. Lee et al. 1990; Caputo et al. 1993, hereafter CDMPQ) show that this holds only for not very blue HB types [(B-R)/(B+V+R) $\leq$+0.8] and that for fixed age and chemical composition the predicted difference $\Delta V$(TO-HB) increases with the HB type, mimicking larger ages in the case of clusters with very blue. horizontal branch morphology.

In reality, as shown in Fig. 2 in CDMPQ, in the case $Z$=0.0001 the constancy of $L$(RR) occurs only in the rather restricted range +0.8 $\geq$ (B-R)/(B+V+R) $\geq$ 0, since even for redder HB type the mean luminosity of RR Lyrae variables seems to increase, leading to overestimate the age of metal-poor clusters with very red HB morphology. Such an effect till now has not been further investigated because very few clusters (some Palomar and Palomar-like clusters) are observed with such low metallicity and red HB morphology. However, similar features are not rare out of the Galaxy. As a fact, several dwarf spheroidal (dSph) galaxies member of the Local Group show [Fe/H]$\leq$-1.8 and (B-R)/(B+V+R)$\leq$ -0.4, and consequently we should correct the observed $\Delta V$(TO-HB) measurement before deriving the cluster age.

However, such systems show also a wide range of $\Delta V$(TO-HB), from the globular-like value of 3.5mag (Ursa Minor) down to 2.5mag (Carina), till the extreme case of Leo I where no HB is observed and the He-burning stars appear as clumping red giants, a feature which suggests an age of very few Gyrs (see e.g. Castellani et al. 1992). Given such a large variation in age, the assumption of $L$(RR) independent of age is no more valid and the great deal of theoretical HB models devoted to Galactic globulars (old and metal-poor) appears to be inadequate since they are based upon He-core mass ($Mc$, in solar unit) and $\Delta Y$ (the amount of extra-helium brought to the surface by the first dredge-up) computed for ages generally in the range of 13 to 18 Gyr. Nor the evolutionary computations aimed at the intermediate-age Galactic open clusters (see e.g. Cassisi et al. 1994) appear suitable since they generally refer to $Z \geq 0.001$.

To cover such a gap, Castellani & Degl'Innocenti (1994, hereafter Paper I) have recently performed a set of computations for masses in the range of 0.6 to 2.0$M\odot$, $Z$=0.0001, 0.0004, and $Y$(MS)=0.23, investigating the evolution from H-burning phase till the onset of He-flash. On this basis, selected isochrones for H-burning stars with age in the range of 15 to 0.7 Gyr were presented and evolutionary zero age horizontal branch (ZAHB) sequences derived for $t$=15, 7, 4, 2, 1, and 0.7, and with masses $MHB$ (in solar unit) in the range of $Mc$ to $MRG$. Since these new models adopt the same original helium, input physics, and computer code as for Straniero & Chieffi (1991, hereafter SC) isochrones ($Z$ in the range of 0.0001 to 0.006 and age from 10 to 20 Gyr) and Castellani et al. (1991, hereafter CCP) HB evolutionary tracks ($Z$ in the range of 0.0001 to 0.02, $t \sim$=15), the theoretical background for $Z$=0.0001, 0.0004 appears now quite complete for extending to lower ages both the analysis of the HB morphology and the $\Delta V$(TO-HB) calibration, providing us with a reliable reference framework for interpreting dSph galaxies and, more generally, any metal-poor stellar system older than 0.7 Gyr.

In this paper we will investigate the connection of the theoretical scenario with observations. The results concerning MS turn-off and ZAHB sequences are discussed in Sect.2, while in Sect.3 we present synthetic models of horizontal branch together with a calibration of $\Delta V$(TO-HB) as a function of age, metallicity and HB morphology. The comparison between theoretical constraints and the relevant observations of dSph galaxies and some Galactic globular clusters is presented in Sect.4, where the main results are also summarized.

## 2. From main sequence to ZAHB

In this section we give the main evolutionary features which are relevant for the present investigation. As a first point, let us recall that in Paper I it has been shown that for age in the range of 0.7 to 15 Gyr the luminosity $\log L$(TO) of the main sequence turn-off is linearly dependent on $\log t$. As shown in Fig.1, the relations given in Paper I are in perfect agreement with SC results and consequently they can be safely used over the whole range $1 \leq t \leq 20$.

Moreover, we find that at each given metallicity the age-dependent values of $Mc$ and $MRG$ fit in harmoniously with the results given by SC for $t \geq 10$ (see Fig.2 and Fig.3). The least squares solutions through the data at $Z=0.0001$ are

$$MRG = 1.705 - 0.999 logt + 0.199[logt]^2 \qquad (1)$$

$$Mc = 0.441 + 0.090 logt - 0.031[logt]^2, \qquad (2)$$

while at $Z=0.0004$ we get

$$MRG = 1.788 - 1.159 logt + 0.269[logt]^2, \qquad (3)$$

$$Mc = 0.464 + 0.041 logt - 0.011[logt]^2. \qquad (4)$$

To discuss the ZAHB sequences, we refer to the results of Paper I, plus an additional sequence at $t=10$ and $Z=0.0001$ which has been computed for the occasion, using the same original helium, computer code and input physics.

As a general feature, at each given age and $MHB$, the ZAHB models with $Z=0.0001$ are systematically bluer than those with $Z=0.0004$, the difference in effective temperature increasing significantly towards larger masses. As a fact, for both metallicities the effective temperature of the ZAHB structure decreases with increasing the mass, reaching the minimum value of $logTe,min=3.74$ ($Z=0.0001$) and 3.72 ($Z=0.0004$) around 1-1.2$M\odot$. After that, more massive structures with $Z=0.0004$ are characterized by higher luminosity (dlog$L$/d$M$ $\sim$0.4) and roughly constant effective temperature (dlog$Te$/d$M$ $\sim$0.05), whereas those with $Z=0.0001$ again by higher luminosity (dlog$L$/d$M$ $\sim$0.4) but with larger effective temperature (dlog$Te$/d$M$ ~0.25). This yields that He-burning massive stars in the range of 1.3$M\odot$ to 1.7$M\odot$ with $Z$=0.0004 should be confined near the red giant branch, whereas those with $Z$=0.0001 have good opportunities to populate an "upper horizontal branch" (UHB). It is worth noticing that a further decrease of the metallicity down to $Z$=0.00001 (see Caloi et al. 1978) would shift the ZAHB "turn-over" to log$Te$ ~3.89, i.e. all the ZAHB models should be very likely confined at the blue side of the instability strip.

Following the variation of $Mc$ with age, the ZAHB luminosity at the mean temperature of the RR Lyrae gap (log$Te$=3.83) cannot be more regarded as independent of the age (see data in Table 1 and Fig. 4). As a matter of example, at $t$=7 and $t$=2 the ZAHB luminosity is $\delta$ log$L$ ~ 0.02 and ~0.09, respectively, lower than at $t$=15.

At $Z$=0.0001 the least-squares solution for the correlation of log$L$(3.83) (as read along the "normal" ZAHB before the turn-over) with age gives

$$logL(3.83) = 1.727 - 0.205/t, \tag{5}$$

which means a negligible variation for $t \geq 15$. Moreover, from the results of the two ZAHBs with $Z$=0.0004, $t$=15 and $t$=7 we derive dlog$L$(3.83)/dlog$Z$=-0.07.

The combination of all the above results provides us with the predicted difference in luminosity between TO and ZAHB as a function of metallicity and age. By transforming luminosity into visual magnitude through Kurucz (1992) bolometric corrections, we derive for each given age the difference $\Delta V$(TO-ZAHB), as listed in Table 2 for $Z$=0.0001. One finds that for ages in the range of 20 to 7Gyr the correlation with log$t$ turns out to be perfectly linear (see Fig. 5), with the least-squares solution through the data (dashed line)

giving

$$logt = -0.082 + 0.374\Delta V(TO - ZAHB). \tag{6}$$

However, as shown in the same Fig.5, this relation could be used over the whole range $1\leq t \leq 20$ since it yields an error of d$t \sim$ -0.4 at $t$=4 and d$t \sim$ +0.3 at $t$=1. We should notice that the hypothesis of constant log$L$(RR), as given f.e. at $t$=15, would yield similar results for $t \geq 7$, while for $t \leq 4$ the estimated ages are systematically smaller by $\sim$ 0.8 Gyr (see dotted line).

As for the $Z$=0.0004 case, at fixed age the predicted difference is larger by $\sim$0.09 mag with respect to $Z$=0.0001. This yields that at fixed $\Delta V$(TO-ZAHB) the resulting age is lower by dlog$t \sim$-0.034.

## 3. Post-ZAHB evolution and synthetic HB

In order to have an immediate insight into the connection between HB evolution and instability strip, let us introduce a "pulsational" HR diagram where the effective temperatures are scaled to the effective temperature ($RE$) at the red edge of the instability strip, as appropriate for $Y$=0.24 and each given mass and luminosity. The location of the red edge, as well as the adopted width of the strip ($\Delta$log$Te$=0.07) are provided from pulsational models computed for masses in the range of 0.58 to 1.5$M\odot$ and luminosity 1.5$\leq$log$L \leq$2.2 (Bono & Stellingwerf 1994, Bono et al. 1994b, Bono & Caputo 1994) which yield the analytical relation

$$RE = 3.889 + 0.048(Y - 0.24) - 0.068 logL + 0.078 log(M/0.65). \tag{7}$$

On this basis, ZAHB sequences and evolutionary tracks for selected stellar masses are presented in Figs.6-8 ($Z$=0.0001) and Figs.9-10 ($Z$=0.0004). All the ZAHB models cover the mass range of $0.70M\odot$ to $MRG$ with intervals $\Delta M$=$0.1M\odot$, except the cases with $Z$=0.0001, $t$=15 (0.70, 0.75, and $0.80M\odot$), $Z$=0.0001, $t$=1 ($0.65M\odot$ is added), and $Z$=0.0004, $t$=15 (0.65, 0.70, and $0.75M\odot$). The evolution of the most massive models ($MHB = MRG$) is from Paper I, that of the remaining models with $t$=15 from CCP, and all the others evolutionary tracks have been computed for the occasion.

For each HB evolutionary track we derive the relative numbers of B,V, and R stars - and then the predicted HB type - by a simple evaluation of the evolutionary time which is spent at the left, within, and at the right of the instability strip. The results are given in Table 3 for $Z$=0.0001 and Table 4 for $Z$=0.0004, together with the estimated difference $\delta V$ between the visual magnitude of the ZAHB at $\log Te$=3.83 and the mean value within the instability strip or, in the case of (B-R)/(B+V+R)=-1, the mean magnitude of the whole HB track, as derived through Kurucz (1992) bolometric corrections.

The results show that at each given metallicity and age the (B-R)/(B+V+R) ratio decreases -as expected - as $MHB$ increases, till only red HB stars are present for masses around $1.0M\odot$. Further increasing $MHB$ leaves the HB stars to the left of the instability strip until ZAHB turn-over and HB evolution push again the He-burning structures into the instability strip, but at higher luminosity than the "normal" RR Lyraes. From the present models we find that the lower total mass for observing such overluminous variables is $\sim 1.3M\odot$ for $Z$=0.0001 and $\sim 1.8M\odot$ for $Z$=0.0004. Moreover, the $2.0M\odot$ ($t$=0.7) models presented in Paper I provide us with further constraints. As a fact, at $Z$=0.0001 the evolutionary track is hotter than the instability strip (HB type $\sim$ 0.99), thus putting for the variables with this metallicity an upper limit of the order of 1.9$M\odot$. On the contrary, the model with $Z$=0.0004 evolves just within the instability strip (HB type $\sim$ -0.03), suggesting masses even moderately larger than 2.0$M\odot$. This kind of overluminous variables has been already correlated with the so-called "anomalous cepheids" by Demarque & Hirshfeld (1975) and Hirshfeld (1980). Their predicted pulsational properties, as well as the comparison with observed data, is matter of a forthcoming paper (Bono et al. 1994a).

Inspection of data in Table 3 discloses the interesting evidence that the HB type is almost univocally determined by the ratio $q$ of the He-core mass to the total mass $MHB$ (see Fig.11), independently of the age. This allows an easy prediction of the (B-R)/(B+V+R) ratio for each assumption on the cluster age (i.e. on $Mc$ and $MRG$, as given by Eqs.1-4) and on the amount of mass loss (i.e. on $MHB$). In particular, for each given age we predict the HB morphology under the hypothesis of $MHB = MRG$ (no mass loss), as given by the solid line in Fig.12. If $t \geq 4$, then the occurrence of mass loss during the red giant phase would yield bluer HB morphologies than the solid line, whereas for lower ages the HB type can vary from +1.0 to -1.0. In other words, the solid line in Fig.12 represents the "permitted" reddest HB type (HB,r) for each given age in the range of 20 to 4Gyr, while at $t \leq 4$ no further constraint besides the obvious limit of HB,r=-1.0 (dashed line) can be given. On this basis, one could use the observed HB morphology as indicator of the upper limit for the cluster age. As a matter of example, a star cluster with $Z$=0.0001 and (B-R)/(B+V+R) =-0.6 or -1.0 cannot be older than 11Gyr or 8Gyr, respectively.

The above analysis for $Z$=0.0001 allows us to interpret the fewer data for $Z$=0.0004

which are shown Fig.13. The effect of the increased metallicity is to shift the HB,r values towards the red so that (B-R)/(B+V+R)=-1.0 occurs already at $t \sim 13$ (see Fig. 14).

The predicted HB,r for the two choices of metallicity are listed in Table 5 as a function of age. It seems worth noticing that a further mild increase of metallicity (f.e. $Z=0.0006$) is sufficient for obtaining HB,r=-1.0 over the whole age range of 1 to 20 Gyr. On the contrary, from Caloi et al. (1978) computations, we roughly estimate that at $Z=0.00001$ redder morphologies than (B-R)/(B+V+R)$\sim$0.7 should be forbidden, whichever is the age.

The data in Table 3 yields a further result worthy of notice: also the mean increase in luminosity $\delta V$ with respect the ZAHB at $\log Te=3.83$ seems to depend on the ratio $q$, independently of the age. This is shown in Fig.15 where the full points refer to evolutionary tracks starting from the "normal" ZAHB, the triangles to tracks populating only the red side of the strip, and the open points to the evolutionary paths starting from the post-turnover ZAHB. The combination of this result with those shown in Fig. 11 and Fig. 13 eventually gives the predicted dependence of $\delta V$ on the (B-R)/(B+V+R) ratio which is shown in Fig.16.

One derives that at $Z=0.0001$ and age in the range of 1 to 20Gyr the difference in magnitude between the ZAHB at $\log Te=3.83$ and the "normal" horizontal branch at the RR Lyrae gap (full points) is quite constant ($\sim$ 0.1mag) for (B-R)/(B+V+R) in the range of +0.8 to -0.3, increasing for redder and bluer HB types. Concerning the purely red HB morphologies (triangles), the minimum difference in magnitude with respect to ZAHB is $\delta V \sim +0.25$mag, while the maximum value depends on the age and amount of mass loss, reaching $\sim$0.7mag at $t=2$-1, $MHB \sim 1.2$. As for the UHB variables, they are characterized by $\delta V$ in the range of 0.8mag ($MHB \sim 1.3$) to 1.3mag ($MHB=1.7$), with an upper limit of $\sim$2mag at $MHB \sim 1.9$.

As for $Z=0.0004$, we derive $\delta V \sim 0.1$mag for (B-R)/(B+V+R) in the range of +0.80 to -0.90. For the purely red HB types, $\delta V$ should start from $\sim 0.2$mag and it may increase up to 1.3mag at $t=1$, $MHB=1.7$. For the more massive models which enter the instability strip during the UHB phase, from the data in Table 4 we estimate that the variables should be brighter by at least $\delta V \sim 1.4$mag ($MHB \sim 1.8$) with respect those populating the normal ZAHB.

Figure 17 and Figure 18 show the predicted difference $\Delta V$(TO-HB) as a function of HB type, for the various ages and the two choices of metallicity. One finds that extremely blue HB clusters may be credited with a larger age by $\delta \log t \sim 0.07$, i.e. $\delta t \sim$ 2-3 Gyr at $t=10$-15, if the observed $\Delta V$(TO-HB) is not systematically corrected according the HB type. Moreover, the figures clearly show that we are not allowed to know the precise age of relatively young ($t \leq 7$) clusters with (B-R)/(B+V+R)=-1.0, owing to the variation of $\delta V$ with the amount of mass-loss. From the estimated maximum and minimum difference $\Delta V$(TO-HB) which are listed in Table 6 for the various ages and the two metallicities, one has that if a difference of , f.e., 2.0 mag is measured between the TO level and the apparent magnitude of a stubby red HB, then the resulting cluster age may range from 2.5 to 4Gyr and from 1.5 to 4Gyr, for $Z=0.0001$ and 0.0004, respectively.

In conclusion, present result from one side confirm that the HB morphology should be considered when dating stellar clusters from the observed difference between turn-off and horizontal branch, and from the other side, provide important information on the role of metallicity and age on some observed parameters such as the width in magnitude of the red giant clumping stars, the difference in magnitude between normal RR Lyrae stars and anomalous cepheids, and the period-luminosity relation of the anomalous cepheids in a given dSph galaxy. Two forthcoming papers (Bono et al. 1994a; Caputo et al. 1994) will deal extensively with these latter subjects.

## 4. Comparison with observed clusters and conclusions

The relevant observational parameters of dSph galaxies and some selected Galactic globular clusters (Ruprecht 106, Palomar 3, and Palomar 4) are listed in Table 7 and Table 8. The [Fe/H] values are an average of current estimates, while the $\Delta V$(TO-HB) difference and HB type are given or measured from the quoted sources.

The maximum permitted ages, as derived from the observed HB morphology and for the two choices of metallicity, are given in columns 4 and 5 of Table 6. As expected from the discussion in Sect.3, these upper limits are strongly dependent on $Z$, increasing for larger metallicity.

The absolute ages, as derived from observed $\Delta V$(TO-HB) and (B-R)/(B+V+R) measurements are listed in Table 7. For the two Palomar clusters, no age is given at $Z=0.0001$ because the two observed parameters are incompatible (see Fig.17). Inspection of the table shows that the age is now very mildly decreasing when increasing the adopted metallicity. The comparison with the age upper limits derived from the HB morphology suggests that on an average the overall metallicity of the red HB clusters here analyzed is closer to $Z=0.0004$ than to $Z=0.0001$.

The main results of the present analysis of H- and He-burning stars with $Z$=0.0001, 0.0004 and age in the range of 1 to 20Gyr can be summarized as follows:

1. The luminosity of ZAHB at the RR Lyrae gap is found to depend on metallicity as well as age as $\log L(3.83)$=1.727-0.07($\log Z$+4.0)-0.205/$t$.

2. The luminosity of the MS turn-off is linearly dependent on $\log t$ and the resulting predicted correlation of $\Delta V$(TO-ZAHB) with metallicity and age can be linearly approximated as $\log t$=-0.082-0.057($\log Z$+4.0)+0.374$\Delta V$(TO-ZAHB).

3. The increase over $L(3.83)$, as the result of post-ZAHB evolution, is a function of metallicity and HB morphology. We derive $\delta V \sim$ 0.1mag for HB types in the range of +0.8 to -0.3 and of +0.8 t0 -0.9 for $Z$=0.0001 and 0.0004, respectively, while for redder and bluer HB types $\delta V$ may increase significantly. Such results should be properly taken into account when the cluster age is derived from the observed difference in magnitude between MS turn-off and horizontal branch (rather than ZAHB).

4. At fixed metallicity and for each given age we evaluate the "permitted" reddest HB morphology. This yields an easy prediction of the maximum age which corresponds with each observed HB morphology. Since the upper limit for a given HB type increases significantly with increasing $Z$, such a procedure appears of particular interest in the case of very metal-poor clusters.

5. The ages which are derived from the difference $\Delta V$(TO-HB) show a mild decrease when increasing the adopted metallicity. As a consequence, the comparison with the age upper limits, as derived from HB morphology, could yield useful constraints to the overall metallicity of the cluster under examination.

## 4. Acknowledgements

We are indebted to V. Castellani for helpful discussions and comments on a draft of this paper.

# Figure captions

Fig. 1: The luminosity of MS turn-off versus age for the two choices of metal content. The solid lines represent the relations given in Paper I. Full points and open points refer to present and SC results, respectively

Fig. 2: As in Fig. 1, but for the mass at the He-flash, for $Z=0.0001$. The solid line is eq.1

Fig. 3: As in Fig.1, but for the mass of He-core, for $Z=0.0001$. The solid line is eq.2

Fig. 4: The ZAHB luminosity at the RR Lyrae gap versus age, for $Z=0.0001$. The solid line is eq.5

Fig. 5: The correlation between age and $\Delta V$(TO-ZAHB), for $Z=0.0001$. The dashed line is eq.6, while the dotted line refers to the case $L$(RR)=const (see text)

Fig. 6: HR location of ZAHB and HB models for the labelled metallicity and age. The vertical lines represent the edges of the instability strip

Fig. 7: As in Fig.6

Fig. 8: As in Fig.6

Fig. 9: As in Fig.6

Fig. 10: As in Fig.6

Fig. 11: HB type versus the ratio $Mc/MHB$, for $Z=0.0001$

Fig. 12: HB type versus age under the hypothesis of no mass-loss (solid line), for $Z=0.0001$

Fig. 13: As in Fig.11, but for $Z=0.0004$

Fig. 14: As in Fig.12, but for $Z=0.0004$

Fig. 15: The difference in visual magnitude $\delta V$ between ZAHB and "normal" variables

(full points), purely red HBs (triangles), and overluminous variables (open points) versus the ratio $Mc/MHB$

Fig. 16: The difference in visual magnitude $\delta V$ versus HB type. Symbols as in Fig.15. The maximum value for the labelled ages is also indicated

Fig. 17: The difference $\Delta V$(TO-HB) as a function of HB type and for the labelled age and metallicity

Fig. 18: As in Fig. 17

Table 1: ZAHB luminosity at LogTe=3.83 as a function of metallicity and age.

|   | Z=0.0001 | | Z=0.0004 | |
|---|---|---|---|---|
| t | $M_c$ | LogL(3.83) | $M_c$ | LogL(3.83) |
| 15 | 0.505 | 1.717 | 0.498 | 1.672 |
| 10 | 0.500 | 1.710 | – | – |
| 7 | 0.495 | 1.698 | 0.490 | 1.656 |
| 4 | 0.480 | 1.662 | – | – |
| 2 | 0.469 | 1.628 | – | – |
| 1 | 0.441 | 1.524 | – | – |

Table 2: Predicted visual magnitude of TO and ZAHB at LogTe=3.83, and the difference $\Delta$V(TO-ZAHB) as a function of age, for Z=0.0001.

| t | Mv(TO) | Mv(ZAHB) | $\Delta$V(TO-ZAHB) |
|---|---|---|---|
| 20 | 4.300 | 0.615 | 3.685 |
| 15 | 3.985 | 0.621 | 3.364 |
| 10 | 3.538 | 0.636 | 2.902 |
| 7 | 3.137 | 0.671 | 2.466 |
| 4 | 2.455 | 0.767 | 1.688 |
| 2 | 2.004 | 0.851 | 1.153 |
| 1 | 1.643 | 1.113 | 0.530 |

Table 3: The fraction of B and V stars, the HB type, and the difference in visual magnitude $\delta V$ for various ages and MHB and for Z=0.0001.

| t | MHB | q | B | V | HB | $\delta V$ |
|---|---|---|---|---|---|---|
| 15 | 0.70 | 0.721 | 0.95 | 0.04 | 0.94 | 0.19 |
| 15 | 0.75 | 0.673 | 0.83 | 0.13 | 0.79 | 0.10 |
| 15 | 0.80 | 0.631 | 0.00 | 0.90 | -0.10 | 0.10 |
| 10 | 0.90 | 0.556 | 0.00 | 0.36 | -0.64 | 0.17 |
| 7 | 1.00 | 0.495 | 0.00 | 0.00 | -1.00 | 0.28 |
| 4 | 0.80 | 0.600 | 0.00 | 0.88 | -0.12 | 0.11 |
| 4 | 1.20 | 0.400 | 0.00 | 0.00 | -1.00 | 0.58 |
| 2 | 0.70 | 0.670 | 0.74 | 0.22 | 0.70 | 0.10 |
| 2 | 0.80 | 0.586 | 0.00 | 0.37 | -0.63 | 0.16 |
| 2 | 0.90 | 0.521 | 0.00 | 0.00 | -1.00 | 0.42 |
| 2 | 1.40 | 0.335 | 0.00 | 0.55 | -0.45 | 1.02 |
| 1 | 1.50 | 0.294 | 0.01 | 0.67 | -0.32 | 1.30 |
| 1 | 1.70 | 0.259 | 0.60 | 0.29 | 0.49 | 1.34 |
| 0.7 | 2.00 | 0.197 | 0.99 | 0.01 | 0.99 | 2.10 |

Table 4: As in table 3, but for Z=0.0004

| t | MHB | q | B | V | HB | $\delta V$ |
|---|---|---|---|---|---|---|
| 15 | 0.65 | 0.766 | 0.98 | 0.02 | 0.97 | 0.33 |
| 15 | 0.70 | 0.711 | 0.43 | 0.51 | 0.37 | 0.08 |
| 15 | 0.75 | 0.664 | 0.00 | 0.27 | -0.73 | 0.08 |
| 7 | 1.00 | 0.490 | 0.00 | 0.00 | -1.00 | 0.39 |
| 2 | 1.40 | 0.341 | 0.00 | 0.00 | -1.00 | 0.87 |
| 1 | 1.70 | 0.275 | 0.00 | 0.00 | -1.00 | 1.33 |
| 0.7 | 2.00 | 0.222 | 0.00 | 0.97 | -0.03 | 2.00 |

Table 5: Permitted reddest HB type as a function of age and metallicity

| t | HB,r (Z=0.0001) | HB,r (Z=0.0004) |
|---|---|---|
| 20 | 0.80 | -0.40 |
| 18 | 0.75 | -0.70 |
| 16 | 0.05 | -0.80 |
| 14 | -0.20 | -0.95 |
| 12 | -0.40 | -1.00 |
| 10 | -0.70 | -1.00 |
| 9 | -0.80 | -1.00 |
| 8 | -1.00 | -1.00 |
| < 8 | -1.00 | -1.00 |

Table 6: Maximum and minimum difference $\Delta V(TO\text{-}HB)$ for purely red HB type as a function of the mass-loss amount (ML), for various ages and for the two metallicities.

| t | $ML_{max}$ | $\Delta V_{min}$ | $ML_{min}$ | $\Delta V_{max}$ |
|---|---|---|---|---|
| | | Z=0.0001 | | |
| 1 | 0.75 | 0.78 | 0.45 | 1.23 |
| 2 | 0.50 | 1.40 | 0.20 | 1.85 |
| 3 | 0.35 | 1.72 | 0.05 | 2.12 |
| 4 | 0.25 | 1.94 | 0.00 | 2.27 |
| 5 | 0.15 | 2.25 | 0.00 | 2.44 |
| 6 | 0.10 | 2.50 | 0.00 | 2.61 |
| 7 | 0.05 | 2.72 | 0.00 | 2.75 |
| 8 | 0.00 | 2.88 | 0.00 | 2.88 |
| | | Z=0.0004 | | |
| 1 | 0.95 | 0.82 | 0.00 | 1.95 |
| 2 | 0.60 | 1.44 | 0.00 | 2.11 |
| 3 | 0.45 | 1.76 | 0.00 | 2.33 |
| 4 | 0.35 | 1.98 | 0.00 | 2.45 |
| 5 | 0.25 | 2.29 | 0.00 | 2.65 |
| 6 | 0.20 | 2.54 | 0.00 | 2.82 |
| 7 | 0.15 | 2.76 | 0.00 | 2.95 |
| 8 | 0.10 | 2.92 | 0.00 | 3.05 |
| 9 | 0.08 | 3.06 | 0.00 | 3.08 |
| 10 | 0.05 | 3.19 | 0.00 | 3.30 |
| 11 | 0.03 | 3.29 | 0.00 | 3.35 |
| 12 | 0.02 | 3.40 | 0.00 | 3.44 |
| 13 | 0.00 | 3.49 | 0.00 | 3.49 |

Table 7: Upper limit for the age of selected globulars and dSph galaxies, as derived from observed HB type and for the two metallicities.

| Name | [Fe/H] | HB | t(Z=0.0001) | t(Z=0.0004) |
|---|---|---|---|---|
| Galaxy | | | | |
| R106 | -1.90 | -0.93 (1) | 8.5 | 14.5 |
| P 3 | -1.78 | -1.00 (2) | 8.0 | 13.0 |
| P 4 | -2.20 | -1.00 (3) | 8.0 | 13.0 |
| dSph galaxies | | | | |
| Ursa | -2.20 | 0.47 (4) | 17.5 | >20 |
| Draco | -2.10 | -0.41 (5) | 12.0 | 20.0 |
| Carina | -1.87 | -0.70 (6) | 10.0 | 18.0 |
| Sculptor | -1.80 | -0.46 (5) | 11.0 | 19.5 |
| Leo I | -2.00 | -1.00 (7) | 8.0 | 13.0 |
| Leo II | -2.00 | -1.00 (8) | 8.0 | 13.0 |
| Sextans | -2.05 | -0.35 (9) | 13.0 | 20.0 |

(1): Buonanno et al. 1993;
(2): Ortolani & Gratton 1989;
(3): Christian & Heasley 1986;
(4): Cudworth et al. 1986;
(5): Kunkel & Demers 1977;
(6): Mould & Aaronson 1983;
(7): Lee et al. 1993;
(8): Demers & Irwin 1993;
(9): Mateo et al. 1991

Table 8: Age of some selected clusters and dSph galaxies, as derived from the observed difference $\Delta$V(TO-HB) and for the two metallicities.

| Name | [Fe/H] | $\Delta$V(TO-HB) | t(Z=0.0001) | t(Z=0.0004) |
|---|---|---|---|---|
| Galaxy | | | | |
| R106 | -1.90 | 3.2 (1) | – | 10.5 |
| P 3 | -1.78 | 3.1 (2) | – | 9.0 |
| P 4 | -2.20 | 3.3 (3) | – | 11.0 |
| dSph galaxies | | | | |
| Ursa | -2.20 | 3.5 (4) | 15.5 | 14.5 |
| Draco | -2.10 | 3.3 (5) | 12.5 | 12.0 |
| Carina | -1.87 | 2.5 (6) | 6.0 | 6.0 |
| Sculptor | -1.80 | 3 (7) | 10.0 | 9.5 |
| Sextans | -2.05 | 3 (8) | 10.0 | 9.5 |

(1): Buonanno et al. 1993;
(2): Ortolani & Gratton 1989;
(3): Christian & Heasley 1986;
(4): Olszewski & Aaronson 1983;
(5): Stetson et al. 1985;
(6): Mould & Aaronson 1983;
(7): Da Costa 1984;
(3): Mateo et al. 1991